\documentclass[a4paper,UKenglish]{lipics}

\graphicspath{{../}}%helpful if your graphic files are in another directory
\title{Topological Data Analysis with Bregman Divergences\footnote{This research is partially supported
               by the {\sc Toposys} project FP7-ICT-318493-STREP.}}
\titlerunning{Topological Data Analysis with Bregman Divergences} %optional, in case that the title is too long; the running title should fit into the top page column

\author[1]{Herbert Edelsbrunner}
\author[2]{Hubert Wagner}
\affil[1]{IST Austria\\
 Am Campus 1, 3400 Klosterneuburg, Austria\\  
\texttt{edels@ist.ac.at}}
\affil[2]{IST Austria\\
  Am Campus 1, 3400 Klosterneuburg, Austria\\
  \texttt{hwagner@ist.ac.at}}
\authorrunning{H. Edelsbrunner and H. Wagner} %mandatory. First: Use abbreviated first/middle names. Second (only in severe cases): Use first author plus 'et. al.'

\Copyright{H. Edelsbrunner and H. Wagner}%mandatory, please use full first names. LIPIcs license is "CC-BY";  http://creativecommons.org/licenses/by/3.0/

\subjclass{F.2.2 Nonnumerical Algorithms and Problems}% mandatory: Please choose ACM 1998 classifications from http://www.acm.org/about/class/ccs98-html . E.g., cite as "F.1.1 Models of Computation". 
\keywords{ Topological data analysis, Bregman divergences,
  persistent homology, proximity complexes, algorithms}% mandatory: Please provide 1-5 keywords
\usepackage{epsfig}
\usepackage{graphicx}
\usepackage{color}
\usepackage{times}
\usepackage{amsmath}
\usepackage{amssymb}
\usepackage{xspace}
\usepackage{pb-diagram}

\newif\ifskipme
\skipmefalse

\makeatletter
\long\def\@makecaption#1#2{%
  \vskip\abovecaptionskip
  \sbox\@tempboxa{\small #1: #2}%
  \ifdim \wd\@tempboxa >\hsize
    \small #1: #2\par
  \else
    \global \@minipagefalse
    \hb@xt@\hsize{\hfil\box\@tempboxa\hfil}%
  \fi
  \vskip\belowcaptionskip}
\makeatother

\newtheorem{result}{}

\newcommand{\eop}{\hfill\usebox{\smallProofsym}\bigskip}  %
\newsavebox{\smallProofsym}                            % smallproofsymbol
\savebox{\smallProofsym}                               %
{%                                                     %
\begin{picture}(6,6)                                   %
\put(0,0){\framebox(6,6){}}                            %
\put(0,2){\framebox(4,4){}}                            %
\end{picture}                                          %
}

\renewcommand{\paragraph}[1]{\vspace{1ex}\noindent{\textbf{#1}}}
\renewcommand{\subparagraph}[1]{\vspace{1ex}\noindent{\textit{#1}}}

\newcommand {\mm}[1] {\ifmmode{#1}\else{\mbox{\(#1\)}}\fi}

\newcommand {\scalprod}[2] {{\langle #1 , #2 \rangle}}
\newcommand {\denselist}{\itemsep 0pt\parsep=1pt\partopsep 0pt}

\newcommand{\Rspace}        {\mm{\mathbb{R}}}
\newcommand{\Xspace}        {\mm{\mathbb{X}}}
\newcommand{\Yspace}        {\mm{\mathbb{Y}}}
\newcommand{\Domain}        {\mm{\Omega}}

\newcommand{\QQQ}           {\mm{\cal Q}}

\newcommand{\DelTri}[2]     {\mm{\rm Del}_{#1}{({#2})}}
\newcommand{\Del}[3]        {\mm{\rm Del}_{{#1}}{({#2};{#3})}}
\newcommand{\Cech}[3]       {\mm{\rm \check{C}ech}_{{#1}}{({#2};{#3})}}
\newcommand{\ECech}[2]       {\mm{\rm \check{C}ech}{({#1};{#2})}}
\newcommand{\Rips}[3]       {\mm{\rm Rips}_{{#1}}{({#2};{#3})}}
\newcommand{\ERips}[2]       {\mm{\rm Rips}{({#1};{#2})}}

\newcommand{\Cradius}[1]    {\mm{\varrho_{#1}^{\rm \check{C}ech}}}
\newcommand{\Dradius}[1]    {\mm{\varrho_{#1}^{\rm Del}}}
\newcommand{\Rradius}[1]    {\mm{\varrho_{#1}^{\rm Rips}}}

\newcommand{\Bfun}          {\mm{{F}}}
\newcommand{\Bdist}[3]      {\mm{{D}_{#1}{({#2}\|{#3})}}}
\newcommand{\BdistFun}[1]   {\mm{{D}_{#1}}}
\newcommand{\Ball}[3]       {\mm{{B}_{{#1}}{({#2};{#3})}}}

\newcommand{\BallP}[3]      {\mm{{B}_{{#1}}'{({#2};{#3})}}}
\newcommand{\BallPNo}       {\mm{{B}'}}
\newcommand{\Vdomain}[2]    {\mm{{V}_{\!#1}{({#2})}}}
\newcommand{\VdomainP}[2]   {\mm{{V}_{#1}'{({#2})}}}

\newcommand{\dime}[1]       {\mm{\rm dim}\,{#1}}

\newcommand{\norm}[1]       {\mm{\|{#1}\|}}
\newcommand{\Edist}[2]      {\mm{\|{#1}-{#2}\|}}

\newcommand{\Skip}[1]       {}

\begin{document}

\maketitle

\begin{abstract}
Given a finite set in a metric space, the topological analysis generalizes hierarchical clustering using a 1-parameter
family of homology groups to quantify connectivity in all dimensions. 
The connectivity is compactly described by the persistence diagram. One limitation of the current framework is the reliance on 
metric distances, whereas in many practical applications objects are compared by non-metric dissimilarity
 measures. 
Examples are the Kullback--Leibler divergence, which is commonly used for comparing text and images,
and the Itakura--Saito divergence, popular for speech and sound. These are two members of the broad family of dissimilarities called Bregman divergences.

We show that the framework of topological data analysis can be extended to general Bregman divergences, widening the scope of possible applications. In particular, we prove that appropriately generalized \v{C}ech and Delaunay (alpha) complexes capture the correct homotopy type, namely that of the corresponding union of Bregman balls. Consequently, their filtrations give the correct persistence diagram, namely the one generated by the uniformly growing Bregman balls. 
Moreover, we show that unlike the metric setting, the filtration of Vietoris-Rips complexes 
may fail to approximate the persistence diagram. We propose algorithms to compute the thus generalized \v{C}ech, Vietoris-Rips and Delaunay complexes and experimentally test their efficiency. 
Lastly, we explain their surprisingly good performance by making a connection with discrete Morse theory.

\end{abstract}

%%% PASTE ME HERE:

\newpage

%\fontsize{11}{13}\selectfont

%%%%%%%%%%%%%%%%%%%%%%%%%%%%%%%%%%%%%%%%%%%%%%%%%%%%%%%%%%%%%%%%%%%%%%%%%%%
%%%%%%%%%%%%%%%%%%%%%%%%%%%%%%%%%%%%%%%%%%%%%%%%%%%%%%%%%%%%%%%%%%%%%%%%%%%
\section{Introduction}
\label{sec1}
%%%%%%%%%%%%%%%%%%%%%%%%%%%%%%%%%%%%%%%%%%%%%%%%%%%%%%%%%%%%%%%%%%%%%%%%%%%
%%%%%%%%%%%%%%%%%%%%%%%%%%%%%%%%%%%%%%%%%%%%%%%%%%%%%%%%%%%%%%%%%%%%%%%%%%%
\setcounter{page}{1}

The starting point for the work reported in this paper is the desire
to extend the basic topological data analysis (TDA) paradigm to data
measured with dissimilarities.
In particular for high-dimensional data, such as discrete
probability distributions, notions of dissimilarity inspired
by information theory behave strikingly different from the Euclidean distance,
which is the usual setting for TDA.
On the practical side, the Euclidean distance is particularly ill-suited
for many types of high-dimensional data; 
see for example \cite{Hua08}, which provides evidence that
the Euclidean distance consistently performs the worst among several
dissimilarity measures across a range of text-retrieval tasks.
A broad class of dissimilarities are the \emph{Bregman divergences}
\cite{Bre67}.
Its most prominent members are
the \emph{Kullback--Leibler divergence} \cite{KuLe51},
which is commonly used both for text documents~\cite{Big03,Hua08}
and for images \cite{DoVe02},
and the \emph{Itakura--Saito divergence} \cite{ItSa68},
which is popular for speech and sound data \cite{FBD09}.
We propose a TDA framework in the setting of Bregman divergences.
Since TDA and more generally computational topology are young and
emerging fields, we provide some context for the reader.
For more a comprehensive introduction, see the recent textbook~\cite{EdHa10}.

\paragraph{Computational topology.}
Computational topology is an algorithmic approach to describing shape in
a coarser sense than computational geometry does.
TDA utilizes such algorithms within data analysis.
One usually works with a finite set of points,
possibly embedded in a high-dimensional space.
Such data may be viewed as a collection of balls of a radius
that depends on the scale of interest.
Intersections reveal the connectivity of the data.
For example, the components of the intersection graph correspond to the
components of the union of balls.

\subparagraph{Homology groups.}
These are studied in the area of algebraic topology, where they are used
to describe and analyze topological spaces; see e.g.\ \cite{Hat02}.
The \emph{connected components} of a space or, dually,
the \emph{gaps} between them are encoded in its
zero-dimensional homology group.
There is a group for each dimension.
For example, the one-dimensional group encodes \emph{loops} or, dually,
the \emph{tunnels},
and the two-dimensional group encodes \emph{closed shells} or, dually,
the \emph{voids}.
Importantly, homology provides a formalism to talk about different kinds
of connectivity and holes of a space that allows for fast algorithms.

\subparagraph{Nerves and simplicial complexes.}
The \emph{nerve} of a collection of balls generalizes the intersection graph
and contains a $k$-dimensional simplex for every $k+1$ balls that
have a non-empty common intersection.
It is a hypergraph that is closed under taking subsets,
a structure known as a \emph{simplicial complex} in topology.
If the balls are convex, then the Nerve Theorem states that this
combinatorial construction captures the topology of the union of balls.
More precisely, the nerve and the union have the same homotopy type
and therefore isomorphic homology groups \cite{Bor48,Ler45}.
This result generalizes to the case in which the balls are not
necessarily convex but their common intersections of all orders
are contractible.
In the context in which we center a ball of some radius
at each point of a given set, the nerve is referred to as the
\emph{\v{C}ech complex} of the points for the given radius. Its $k$-skeleton
is obtained by discarding simplices of dimension greater than $k$.
The practice-oriented reader will spot a flaw in this setup:
fixing the radius is a serious drawback that limits data analysis applications.

\subparagraph{Persistent homology.}
To remedy this deficiency, we study the evolution of the topology
\emph{across all scales}, thus developing what we refer to as
\emph{persistent homology}.
For graphs and connected components, this idea is natural
but more difficult to flesh out in full generality. 
In essence, one varies the radius of balls from $0$ to $\infty$,
giving rise to a nested sequence of spaces, called a \emph{filtration}. Topological features, namely homology classes of different dimensions,
are created and destroyed along the way. 
In practice, one computes the \emph{persistence diagram} of a filtration,
which discriminates topological features based on their lifetime,
or \emph{persistence}. The persistence diagram serves as a compact \emph{topological descriptor} of a dataset,
which is provably robust against noise.
Owing to its algebraic and topological foundations, the theory is very general.
Importantly, the Nerve Theorem extends to filtrations~\cite[Lemma 3.4]{ChOu08}, 
so we can often restrict our considerations to complexes for a fixed radius. 
Moreover, the existing algorithms for persistence diagrams can be used
without modification.

\subparagraph{TDA in the Bregman setting.}
In the light of the above, there are only two obstacles to applying
topological data analysis to data measured with Bregman divergences. 
We need to prove that the Nerve Theorem applies also when the balls
are induced by Bregman divergences, and we need to provide efficient algorithms
to construct the relevant complexes.
The main complication is that the balls may be nonconvex,
which we overcome by combining results from convex analysis and topology.

\subparagraph{Applications.}
Persistence is an important method within TDA,
which has been successfully used in a variety of applications.
In low dimensions, it was for example used to shed light on the distribution of
matter in the Universe \cite{Sou11} and to characterize the structure
of atomic configurations in silica glass \cite{NHHEN15}.
As for high-dimensional data, Chan \emph{et al}.\ analyze viral DNA
and relate persistent cycles with recombinations \cite{CCR13},
and Port \emph{et al}.\ study languages leaving the interpretation
of a persistent cycle in the Indo-Germanic family open \cite{PGGC15}.

\paragraph{Related work.}
This paper is the first work at the intersection of topology
and Bregman divergences.
We list related papers in relevant fields.
In machine learning, Banerjee \emph{et al}.\ use the family
of Bregman divergences as the unifying framework for
clustering algorithms~\cite{BMDG05}.
The field of information geometry deals with selected Bregman divergences
and related concepts from a geometric perspective~\cite{AmNa00}. 
Building on the classical work of Rockafellar~\cite{Roc70} in convex analysis,
Bauschke and Borwein are the first to use the Legendre transform
for analyzing Bregman divergences~\cite{BaBo97}.
Boissonnat, Nielsen and Nock \cite{BNN10} use similar methods to make 
significant contributions at the intersection of computational geometry 
and Bregman divergences.
In particular, they study the geometry of Bregman balls and
Delaunay triangulations, but not the topologically more interesting Delaunay, or \emph{alpha}, complexes. In the Euclidean setting, the basic constructions are well understood~\cite{BaEd16, Zom10},
including approximations, which are interesting and useful, but beyond the scope of this paper.
%As we are interested in the topology of systems of Bregman balls,
%their work is most directly related to ours. 

%There is a paper on constructing Cech complexes: Efficient construction of the Cech complex, Stefan Dantchev,, Ioannis Ivrissimtzis

\paragraph{Results.}
This paper provides the first general TDA framework that applies to
high-dimensional data measured with non-metric dissimilarities.
Indeed, prior high-dimensional applications of TDA were restricted
to low-dimensional homology, required custom-made topological results,
or used common metrics such as the Euclidean and the Hamming distances,
which are often not good choices for such data.
We list the main technical contributions:
\begin{enumerate}\denselist
  \item\label{res:contr}  We show that the balls under any Bregman divergence
    have common intersections that are either empty of contractible.      
  \item\label{res:gap}    We show that the persistence diagram
    of the Vietoris--Rips complex can be arbitrarily far from that
    of the filtration of the union on Bregman balls.
  \item\label{res:morse}  We show that the radius functions that correspond
    to the \v{C}ech and Delaunay complexes for Bregman divergences are generalized discrete
    Morse functions.
  \item\label{res:algs} We develop algorithms for computing \v{C}ech and Delaunay radius functions for Bregman divergences,
 which owe their speed to non-trivial structural properties implied by Result~\ref{res:morse}.
  
\end{enumerate}  
Most fundamental of the four is Result \ref{res:contr}, which forms the
theoretical foundation of TDA in the Bregman setting.
It implies that the \v{C}ech and Delaunay complexes for a given
radius have the same homotopy type as the union of Bregman balls.
Combined with the Nerve Theorem for filtrations, it also implies
that the filtration of \v{C}ech and Delaunay complexes
have the same persistence diagram as the filtration of the unions.
In the practice of TDA, the filtration of Vietoris--Rips complexes
is often substituted for the filtration of \v{C}ech or Delaunay complexes.
For metrics, this is justified by the small bottleneck distance between
the persistence diagrams if drawn in log-log scale.
Result \ref{res:gap} shows that such a substitution is not generally
justified for Bregman divergences.
In other words, for some Bregman divergences higher order interactions
have to be taken into account explicitly as they are not approximated
by implications of pairwise interactions.
To appreciate Results \ref{res:morse} and \ref{res:algs}, we note
that the \emph{\v{C}ech radius function} maps every simplex to the
smallest radius, $r$, such that the simplex belongs to the \v{C}ech complex
for radius $r$, and similarly for Delaunay.
Being a generalized discrete Morse function 
has important structural
consequences that make it possible to construct
\v{C}ech and Delaunay complexes in an output-sensitive manner. We support this claim
with experiments.

\subparagraph{Implications.}
Our results open up a new area of research at the intersection
of geometry, topology, algorithms and data analysis.
On the application side, it enables TDA for a wide variety of data.
Moreover, it connects topology with information theory and statistics,
where Bregman divergences play a significant role.
Finally, efficient algorithms and data structures are needed to handle
large datasets. Considerable progress has been made within the TDA community,
but we believe a collaboration with the wider computer science community
would be fruitful.

%\subparagraph{Techniques.}
%We use tools from computational geometry, topology, and convex analysis.
%The main result is derived using the Legendre transform
%applied to strictly convex functions.

\subparagraph{Scope.}
The aim of this paper is to show that the machinery of persistent homology
is applicable to different kinds of high-dimensional data.
While questions remain, we provide a solid foundation for further developments.

\paragraph{Outline.}
Section \ref{sec2} reviews the concept of Bregman divergences,
including an elementary description of the Legendre transform.
Section \ref{sec3} proves the contractibility of common
intersections of Bregman balls and introduces
\v{C}ech, Delaunay, and Vietoris--Rips complexes in the Bregman setting.
Section \ref{sec4} introduces the \v{C}ech and Delaunay radius functions
and explains algorithms for constructing them.
Section \ref{sec5} concludes this paper.

%\clearpage
%%%%%%%%%%%%%%%%%%%%%%%%%%%%%%%%%%%%%%%%%%%%%%%%%%%%%%%%%%%%%%%%%%%%%%%%%%%
%%%%%%%%%%%%%%%%%%%%%%%%%%%%%%%%%%%%%%%%%%%%%%%%%%%%%%%%%%%%%%%%%%%%%%%%%%%
\section{Bregman Divergences}
\label{sec2}
%%%%%%%%%%%%%%%%%%%%%%%%%%%%%%%%%%%%%%%%%%%%%%%%%%%%%%%%%%%%%%%%%%%%%%%%%%%
%%%%%%%%%%%%%%%%%%%%%%%%%%%%%%%%%%%%%%%%%%%%%%%%%%%%%%%%%%%%%%%%%%%%%%%%%%%

Bregman divergences are sometimes called \emph{distances}
because they measure dissimilarity.
As we will see shortly, they are generally not symmetric,
and they always violate the triangle inequality.
So really they satisfy only the first axiom of a metric,
mapping ordered pairs to non-negative numbers and to zero iff
the two elements are equal. 

We begin with a formal introduction of the concept, which originated in the paper by Bregman~\cite{Bre67}.
Their basic properties are well known; see the recent paper by Boissonnat, Nielsen and Nock~\cite{BNN10}. 
We stress that our setting is slightly different: following Bauschke and Borwein~\cite{BaBo97}, we define the divergences in terms of \emph{functions of Legendre type}. The crucial benefit of this additional requirement is that the conjugate of a function of Legendre type is again a function of Legendre type, even if the domain is bounded as in the important case of the standard simplex. In contract, the conjugate of a differentiable and strictly convex function that is not of Legendre type is not necessarily again a convex function.

\begin{figure}[hbt]
  \hspace{-0.2in}
  \centering {\includegraphics[width=3.4in]{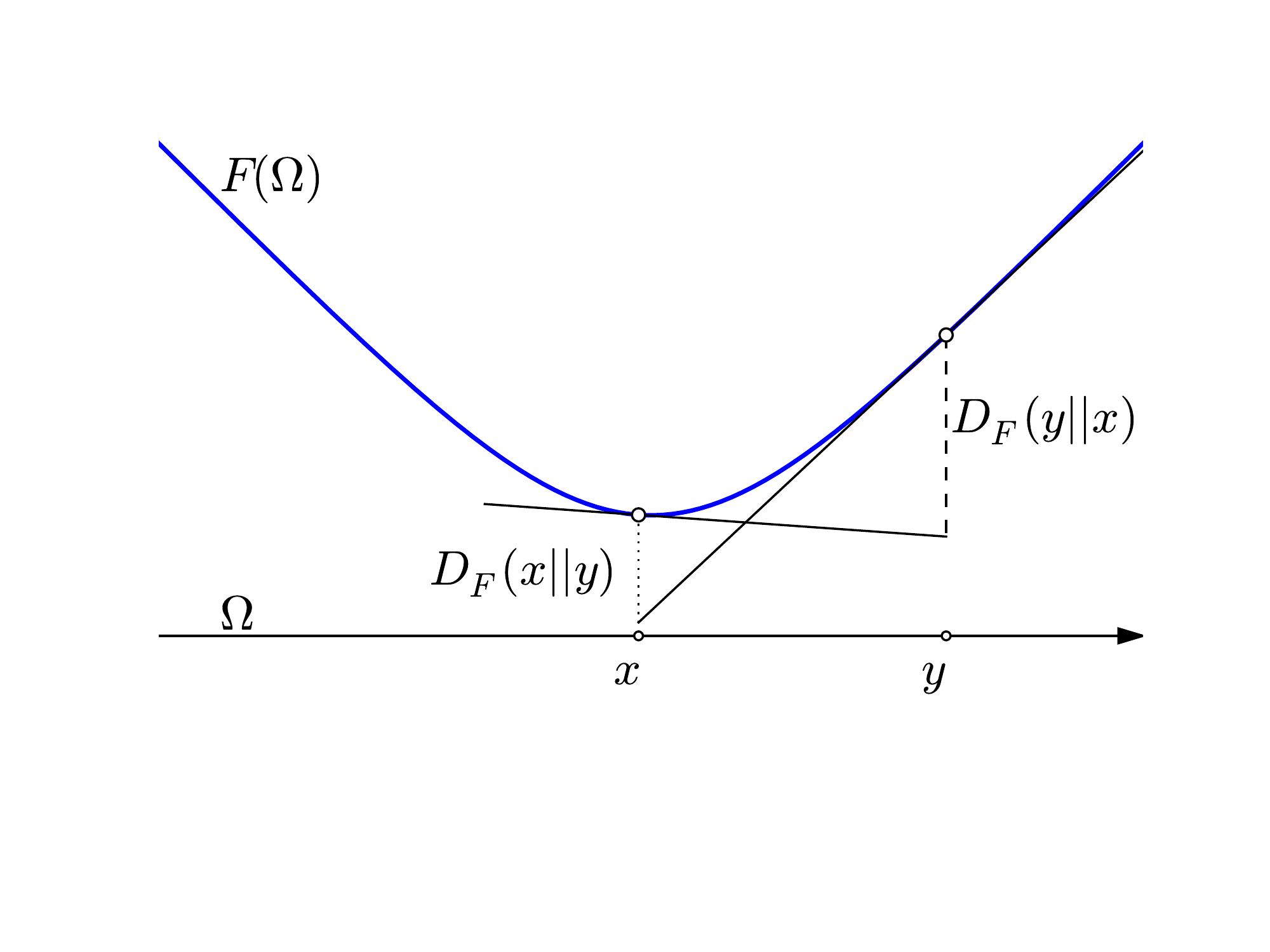}}
  \vspace{-0.6in}	
  \caption{Geometric interpretation of the Bregman divergence
    associated with the function $\Bfun$ on $\Domain$.}
 \label{fig:Bdistance}
\end{figure}

\begin{figure}[hbt]
 \hspace{-0.2in}
 \centering {\includegraphics[width=3.4in]{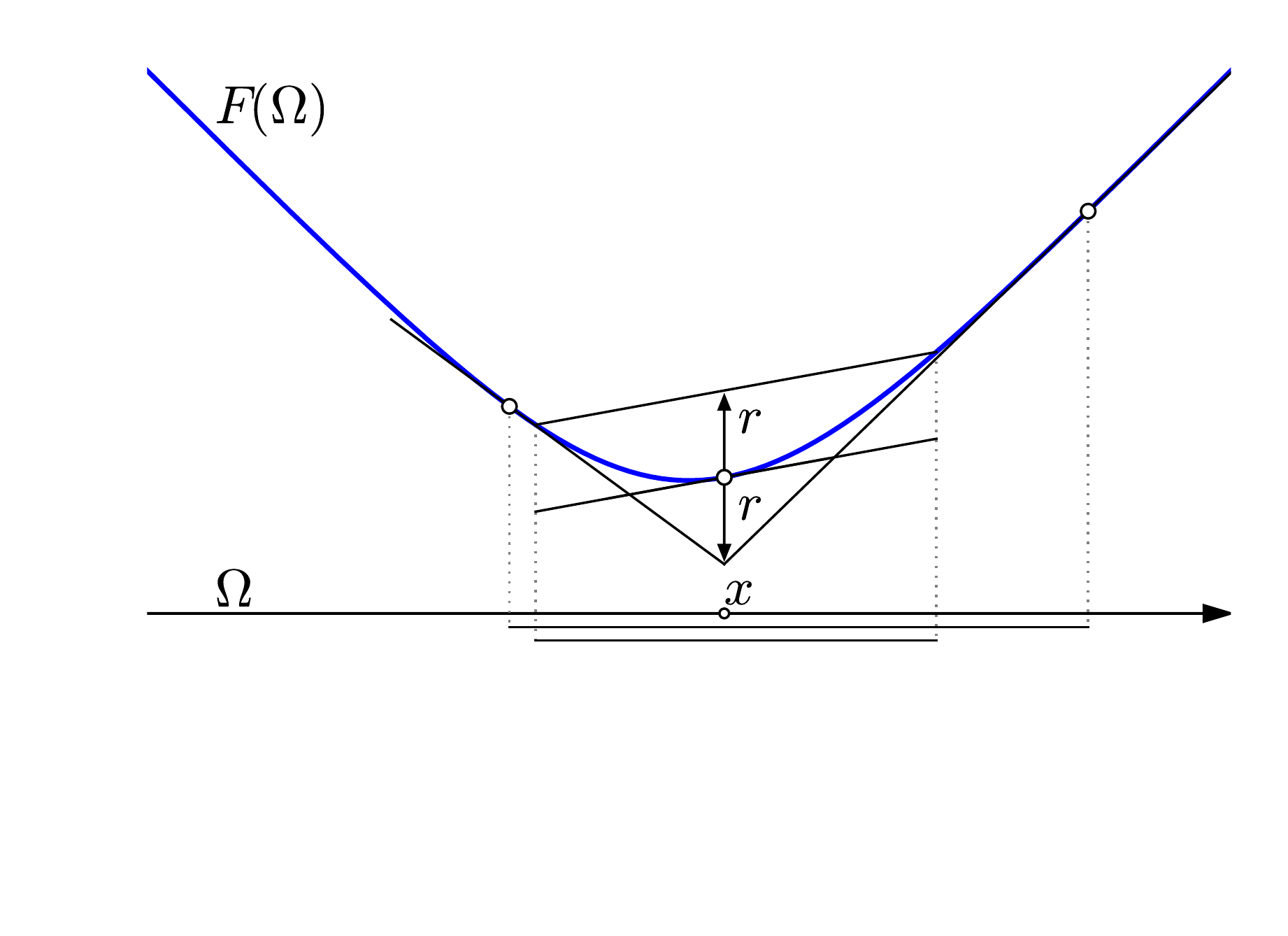}}
 \vspace{-0.6in}
 \caption{The primal Bregman ball with center $x$ is obtained by
   illuminating the graph of $\Bfun$ from below.
   In contrast, the dual Bregman ball is constructed by cutting
   the graph with the elevated line.}
 \label{fig:Bball}
\end{figure}

\begin{figure}[hbt]
 \centering {\includegraphics[width=2.8in]{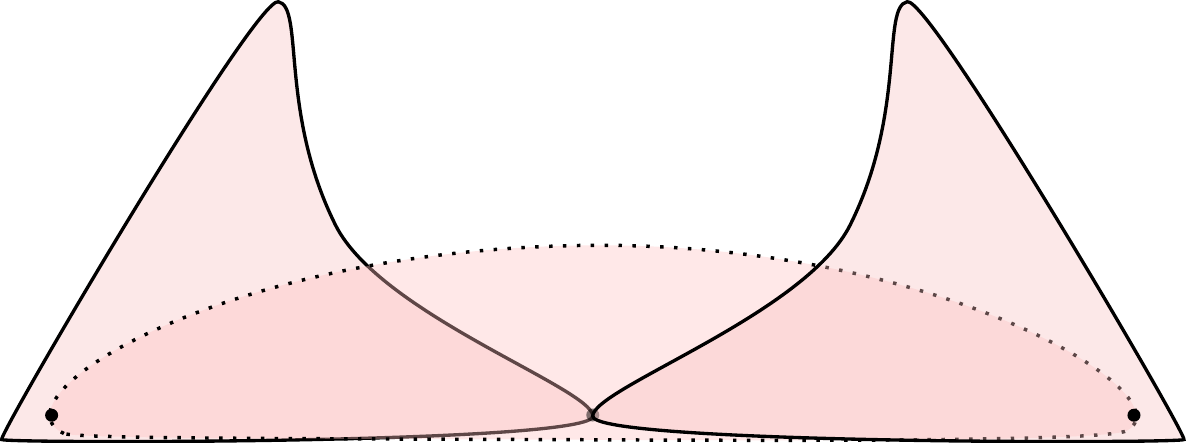}}
 \caption{Two primal Itakura--Saito balls and the dual Itakura--Saito ball
  centered at the point where the primal balls touch.
   Its boundary passes through the centers of the primal balls.}
 \label{fig:itsaballs2}
\end{figure}

%%%%%%%%%%%%%%%%%%%%%%%%%%%%%%%%%%%%%%%%%%%%%%%%%%%%%%%%%%%%%%%%%%%%%%%%%%%
\paragraph{Functions of Legendre type.}
Let $\Domain \subseteq \Rspace^n$ be an nonempty open convex set
and $\Bfun \colon \Domain \to \Rspace$ a strictly convex differentiable function. In addition, we require that  the length of the gradient of $\Bfun$ goes to infinity whenever we approach
the boundary of $\Domain$. Following \cite[page 259]{Roc70}, we say that $\Bfun \colon \Domain \to \Rspace$ is a function of \emph{Legendre type}. As suggested by the naming convention, these conditions are crucial
when we apply the Legendre transform to $\Bfun$.
The last condition prevents us from arbitrarily restricting the domain
and is vacuous whenever $\Domain$ does not have a boundary,
for example when $\Domain = \Rspace^n$.
For points $x,y \in \Domain$, the \emph{Bregman divergence}
from $x$ to $y$ associated with $\Bfun$ is 
the difference between $\Bfun$
and the best linear approximation of $\Bfun$ at $y$,
both evaluated at $x$:
\begin{align}
  \Bdist{\Bfun}{x}{y} &= \Bfun(x)
      - \left[\Bfun(y) + \scalprod{\nabla \Bfun(y)}{x-y}\right] .
  \label{eqn:Bdistance}
\end{align}
As illustrated in Figure \ref{fig:Bdistance},
we get $\Bdist{\Bfun}{x}{y}$ by first drawing the hyperplane that
touches the graph of $\Bfun$
at the point $(y, \Bfun(y))$. We then intersect the vertical line
that passes through $x$ with the graph of $\Bfun$ and the said hyperplane:
the Bregman divergence is the height difference between the two intersections.
Note that it is not necessarily symmetric:
$\Bdist{\Bfun}{x}{y} \neq \Bdist{\Bfun}{y}{x}$ for most $\Bfun, x, y$.
Accordingly, we introduce two balls of radius $r \geq 0$ centered
at a point $x \in \Domain$:
the \emph{primal Bregman ball} containing all points $y$ so that
the divergence from $x$ to $y$ is at most $r$,
and the \emph{dual Bregman ball} containing all points $y$ so that
the divergence from $y$ to $x$ as at most $r$:
\begin{align}
  \Ball{\Bfun}{x}{r}   &=  \{ y \in \Domain \mid \Bdist{\Bfun}{x}{y} \leq r \};
    \label{eqn:ball1} \\
  \BallP{\Bfun}{x}{r}  &=  \{ y \in \Domain \mid \Bdist{\Bfun}{y}{x} \leq r \}.
    \label{eqn:ball2}
\end{align}
To construct the primal ball geometrically,
we take the point $(x, \Bfun(x)-r)$ at height $r$ below
the graph of $\Bfun$ and shine light along straight half-lines emanating
from this point onto the graph.
The ball is the vertical projection of the illuminated
portion onto $\Rspace^n$; see Figure \ref{fig:Bball}.
To construct the dual ball geometrically,
we start with the hyperplane that touches the graph of $\Bfun$
at $(x, \Bfun(x))$, translating it to height $r$ above the initial position.
The ball is the vertical projection of the portion of the graph
below the translated hyperplane onto $\Rspace^n$;
see again Figure \ref{fig:Bball}.
Since $\BdistFun{\Bfun}$ is not necessarily symmetric,
the two Bregman balls are not necessarily the same.
Indeed, the dual ball is necessarily convex while the primal ball is not.
\begin{result}[Convexity Property]
  $\BdistFun{\Bfun} \colon \Domain \times \Domain \to \Rspace$
  is strictly convex in the first argument but not
  necessarily convex in the second argument.
\end{result}
\proof
  Fixing $y$, set $f(x) = \Bdist{\Bfun}{x}{y}$.
  According to \eqref{eqn:Bdistance},
  $f$ is the difference between $\Bfun$ and an affine function;
  compare with the geometric interpretation of the dual Bregman ball.
  The strict convexity of $\Bfun$ implies the strict convexity of $f$.
  This argument does not apply to $g(y) = \Bdist{\Bfun}{x}{y}$,
  which we obtain by fixing $x$,
  and it is easy to find an example in which $g$ is non-convex;
  see Figure \ref{fig:itsaballs2}.
\eop

%%%%%%%%%%%%%%%%%%%%%%%%%%%%%%%%%%%%%%%%%%%%%%%%%%%%%%%%%%%%%%%%%%%%%%%%%%%
\paragraph{Legendre transform and conjugate function.}
In a nutshell, the Legendre transform applies elementary polarity
to the graph of $\Bfun$, giving rise to the graph of another,
conjugate function, $\Bfun^* \colon \Domain^* \to \Rspace$,
that relates to $\Bfun$ in interesting ways. 
If $\Bfun$ is of Legendre type then so is $\Bfun^*$;
see \cite[Theorem 26.5]{Roc70}.

\begin{figure}[hbt]
 \hspace{-0.3in}
 \centering {\includegraphics[width=3.4in]{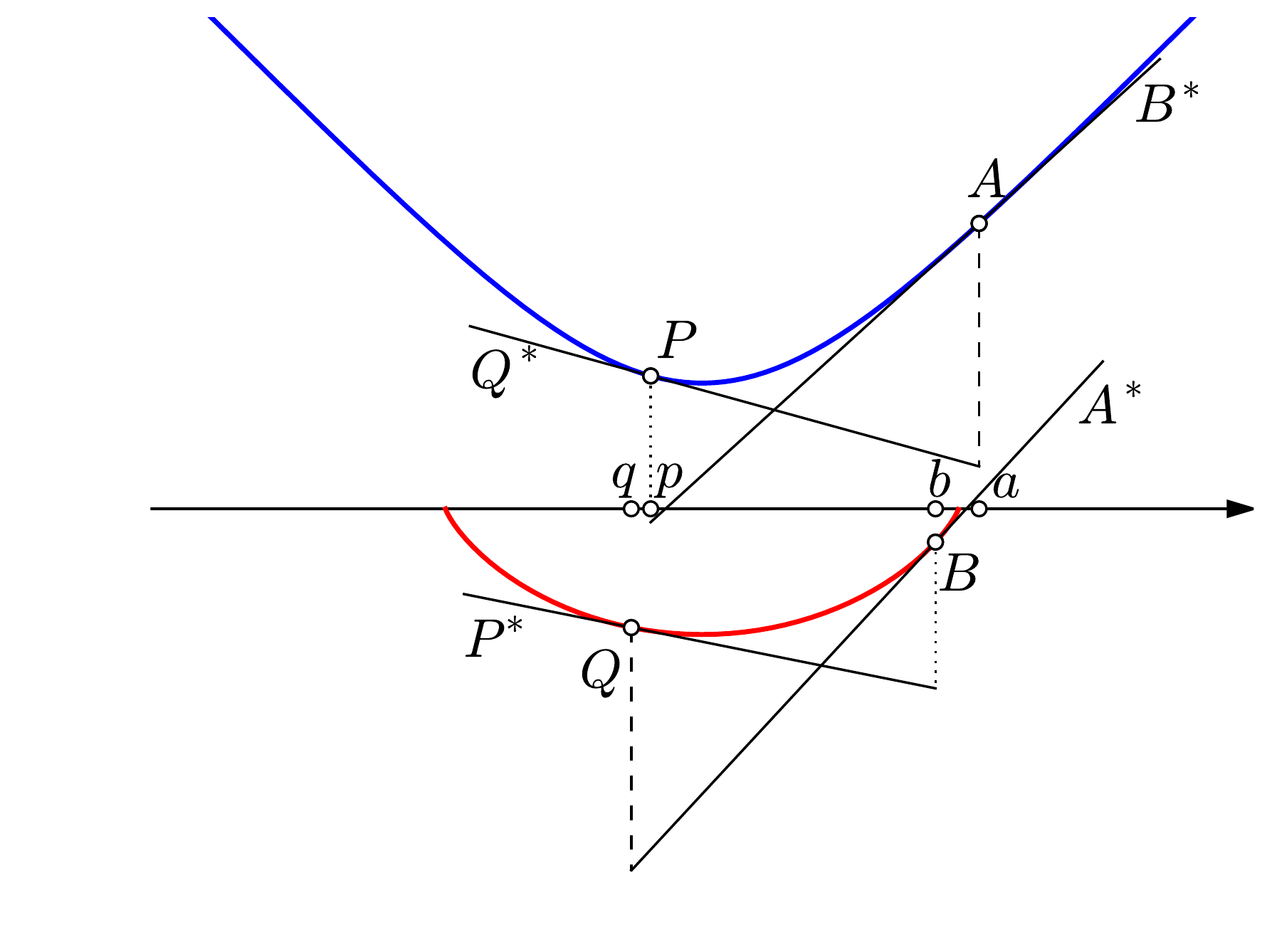}}
 \vspace{-0.2in}
 \caption{\emph{Top}:  the graph of $\Bfun$ and the tangent lines
   that illustrate the two Bregman divergences between $a$ and $p$
   associated with $\Bfun$.
   \emph{Bottom}:  the graph of $\Bfun^*$ and the tangent lines
   that illustrate the two Bregman divergences between $b = a^*$ and $q = p^*$
   associated with $\Bfun^*$.}
 \label{fig:conjugate}
\end{figure}

The notion of polarity we use in this paper relates points
in $\Rspace^n \times \Rspace$ with affine functions $\Rspace^n \to \Rspace$.
Specifically, it maps a point $C = (c, \gamma)$ to the function
defined by $C^* (x) = \scalprod{c}{x} - \gamma$,
and it maps $C^*$ back to $(C^*)^* = C$.
We refer to Figure \ref{fig:conjugate} for an illustration
and to Appendix \ref{appA} for more details.
As a first step in constructing the conjugate function, we get $\Domain^*$
as the set of points $e = c^* = \nabla \Bfun(c)$ with $c \in \Domain$.
We define $h \colon \Domain \to \Domain^*$ by mapping $c$ to $h(c) = c^*$.
Note that differentiability of strictly convex functions implies
\emph{continuous} differentiability~\cite[Theorem 2.86]{DhDu12}, hence $h$
is a homeomorphism between the two domains.

The \emph{conjugate function}, $\Bfun^* \colon \Domain^* \to \Rspace$,
is then defined by mapping $e$ to $\Bfun^* (e) = \epsilon$
such that $(e, \epsilon)$ is the polar point of the affine function
whose graph touches the graph of $\Bfun$ in the point $(c, \Bfun(c))$.
Writing $b = a^*$ and $q = p^*$, we get
%% \eqref{eqn:dual3} and \eqref{eqn:dual4} and get
\begin{align}
  \Bdist{\Bfun^*}{b}{q}  &=  \Bfun^*(b) - P^*(b)  \geq  0 ,
    \label{eqn:nonnegative3} \\
  \Bdist{\Bfun^*}{q}{b}  &=  \Bfun^*(q) - A^*(q)  \geq  0 
    \label{eqn:nonnegative4}
\end{align}
from \eqref{eqn:nonnegative1}, \eqref{eqn:nonnegative1}
and \eqref{eqn:dual3}, \eqref{eqn:dual4} in Appendix \ref{appA};
see again Figure \ref{fig:conjugate}.
The left-hand sides of \eqref{eqn:nonnegative3} and \eqref{eqn:nonnegative4}
are both non-negative and vanish iff $b = q$.
Since this is true for all points $b, q \in \Domain^*$,
$\Bfun^*$ is strictly convex, provided $\Domain^*$ is convex. Proving that this assumption is always fulfilled is more involved. We therefore resort to a classical theorem~\cite[Theorem 26.5]{Roc70}, which states that $\Bfun^*$ is again of Legendre type and, in particular, $\Domain^*$ is convex. 
Hence, $\Bfun^*$ defines a Bregman divergence and, importantly,
this divergence is symmetric to the one defined by $\Bfun$.
\begin{result}[Duality Property]
  Let $\Bfun \colon \Domain \to \Rspace$
  and $\Bfun^* \colon \Domain^* \to \Rspace$
  be conjugate functions of Legendre type.
  Then $\Bdist{\Bfun}{a}{p} = \Bdist{\Bfun^*}{p^*}{a^*}$
  for all $a, p \in \Domain$.
\end{result}

\ifskipme
\proof
  Let $A = (a, \Bfun(a))$ and $P = (p, \Bfun(p))$ be two points on the graph of $\Bfun$,
  and let $B^* (x) = \Bfun(a) + \scalprod{\nabla \Bfun(a)}{x-a}$ and
          $Q^* (x) = \Bfun(p) + \scalprod{\nabla \Bfun(p)}{x-p}$
  be the corresponding affine functions.
  Then $\Bdist{\Bfun}{a}{p} = \Bfun(a) - Q^* (a)$,
  namely the height difference between the graphs
  of $\Bfun$ and $Q^*$ at $a$.

  Let $b = \nabla \Bfun(a)$, $q = \nabla \Bfun(p)$,
  and $\beta, \psi \in \Rspace$
  such that $B = (b, \beta)$ and $Q = (q, \psi)$.
  By construction, $\beta = \Bfun^* (b)$ and $\psi = \Bfun^* (q)$.
  The conjugate Bregman divergence is
  $\Bdist{\Bfun^*}{q}{b} = \Bfun^* (q) - A^* (q)$,
  and applying \eqref{eqn:duality}
  with $C = Q$ and $S = A$, we get
  $\Bfun(a) - Q^*(a) = \Bfun^* (q) - A^* (q)$.
  The claimed relation follows.
\eop
\fi

In words, the Legendre transform preserves the divergences,
but it does so by exchanging the arguments.
This is interesting because $\BdistFun{\Bfun}$
is strictly convex in the first argument
and so is $\BdistFun{\Bfun^*}$, only that its first argument corresponds to the
second argument of $\BdistFun{\Bfun}$.
To avoid potential confusion, we thus consider the primal and dual
Bregman balls of $\Bfun^*$:
\begin{align}
  \Ball{\Bfun^*}{u}{r}  &=  \{ v \in \Domain^* \mid \Bdist{\Bfun^*}{u}{v} \leq r \} ,
    \label{eqn:ball3} \\
  \BallP{\Bfun^*}{u}{r} &=  \{ v \in \Domain^* \mid \Bdist{\Bfun^*}{v}{u} \leq r \} ,
    \label{eqn:ball4}
\end{align}
where we write $u = x^*$ and $v = y^*$ so we can compare the two
balls with the ones defined in \eqref{eqn:ball1} and \eqref{eqn:ball2}.
As mentioned earlier, both dual balls are necessarily convex
while both primal balls are possibly non-convex.
Recall the homeomorphism $h \colon \Domain \to \Domain^*$ that
maps $x$ to $x^*$.
It also maps $\Ball{\Bfun}{x}{r}$ to $\BallP{\Bfun^*}{u}{r}$
and $\BallP{\Bfun}{x}{r}$ to $\Ball{\Bfun^*}{u}{r}$.
In words, it makes the non-convex ball convex and the convex ball non-convex,
and it does this while preserving the divergences.
We use this property to explain the necessity on using functions of Legendre type; it also plays a crucial role later.
Consider a dual Bregman ball with a non-convex conjugate image, namely the corresponding primal ball.
Then the restriction of $\Bfun$ to this dual ball is strictly convex and differentiable. However,
it is not of Legendre type and its conjugate has a non-convex domain.
%%%%%%%%%%%%%%%%%%%%%%%%%%%%%%%%%%%%%%%%%%%%%%%%%%%%%%%%%%%%%%%%%%%%%%%%%%%

\paragraph{Examples.}
We close this section with a short list of functions,
their conjugates, and the corresponding Bregman divergences.
\emph{Half the squared Euclidean norm} maps a point $x \in \Rspace^n$
to $\Bfun(x) = \tfrac{1}{2} \norm{x}^2$.
The gradient is $\nabla \Bfun(x) = x$,
and the conjugate is defined by $\Bfun^* (x) = \Bfun(x)$.
The divergence associated with $\Bfun$ is
\emph{half the squared Euclidean distance}:
\begin{align}
  \Bdist{\Bfun}{x}{y}  &=  \tfrac{1}{2} \Edist{x}{y}^2 .
\end{align}
This Bregman divergence is special because it is symmetric in the
two arguments.

The \emph{Shannon entropy} of a discrete probability distribution
is $- \sum_{i=1}^n x_i \ln x_i$.
To turn this into a convex function, we change the sign,
and to simplify the computations,
we subtract the sum of the $x_i$, defining
$\Bfun(x) = \sum_{i=1}^n [x_i \ln x_i - x_i]$
over the positive orthant, which we denote as $\Rspace^n_+$.
The gradient is $\nabla \Bfun (x) = [\ln x_1, \ln x_2, \ldots, \ln x_n]^T$,
and the conjugate is the \emph{exponential function},
$\Bfun^* (u) = \sum_{i=1}^n e^{u_i}$, with $u = x^*$,
defined on $\Rspace^n$.
Associated with $\Bfun$ is the \emph{Kullback--Leibler divergence}
and with $\Bfun^*$ is the \emph{exponential loss}:
\begin{align}
  \Bdist{\Bfun}{x}{y} &= \sum_{i=1}^ n \left[x_i \ln \tfrac{x_i}{y_i}-x_i+y_i\right], \\
  \Bdist{\Bfun^*}{u}{v} &= \sum_{i=1}^ n \left[e^{u_i} - (u_i-v_i+1) e^{v_i} \right].
\end{align}
The Kullback--Leibler is perhaps the best known Bregman divergence;
it is also referred to as the \emph{information divergence},
\emph{information gain}, \emph{relative entropy};
see \cite[page 57]{AmNa00}.
If applied to finite distributions, $\Bfun$ would be defined on the
standard $(n-1)$-simplex, where it measures the difference in information
when we go from $y$ to $x$.
It also measures the expected number of extra bits required to code
samples from $x$ using a code that is optimized for $y$ instead of for $x$. 
Since the $(n-1)$-simplex is the intersection of $\Rspace^n_+$ with a
 hyperplane, this restriction of $\Bfun$ is again of Legendre type. 
In this particular case, we can extend the function to the \emph{closed} 
$(n-1)$-simplex, so that some coordinates may be zero, 
provided we accept infinite divergences for some pairs. 
In other words, the framework is also suitable for sparse data, 
pervasive for example in text-retrieval applications.

The \emph{Burg entropy} maps a point $x \in \Rspace^n_+$ to
$\Bfun(x) = \sum_{i=1}^n [1 - \ln x_i]$.
The components of the gradient are $- 1 / x_i$, for $1 \leq i \leq n$.
The conjugate is the function $\Bfun^* \colon \Rspace^n_- \to \Rspace$
defined by $\Bfun^* (u) = \sum_{i=1}^n \left[ 1 - \ln |u_i| \right]$.
Associated with $\Bfun$ is the \emph{Itakura--Saito divergence}:
\begin{align}
  \Bdist{\Bfun}{x}{y}  &=  \sum_{i=1}^n \left[ \tfrac{x_i}{y_i}
                               - \ln \tfrac{x_i}{y_i} - 1 \right] .
\end{align}
We note that $\Bfun$ and $\Bfun^*$ are very similar, but their domains
are diagonally opposite orthants.
Indeed, the Itakura--Saito distance is not symmetric and generates
non-convex primal balls; see Figure \ref{fig:itsaballs2}.

%\clearpage
%%%%%%%%%%%%%%%%%%%%%%%%%%%%%%%%%%%%%%%%%%%%%%%%%%%%%%%%%%%%%%%%%%%%%%%%%%%
%%%%%%%%%%%%%%%%%%%%%%%%%%%%%%%%%%%%%%%%%%%%%%%%%%%%%%%%%%%%%%%%%%%%%%%%%%%
\section{Proximity Complexes for Bregman divergences}
\label{sec3}
%%%%%%%%%%%%%%%%%%%%%%%%%%%%%%%%%%%%%%%%%%%%%%%%%%%%%%%%%%%%%%%%%%%%%%%%%%%
%%%%%%%%%%%%%%%%%%%%%%%%%%%%%%%%%%%%%%%%%%%%%%%%%%%%%%%%%%%%%%%%%%%%%%%%%%%

In this section, we extend the standard constructions of topological 
data analysis (\v{C}ech, Vietoris--Rips, Delaunay complexes)
to the setting of Bregman divergences. 
Importantly, we prove the contractibility of non-empty
common intersections of Bregman balls and Voronoi domains.
This property guarantees that the \v{C}ech and Delaunay
complexes capture the correct homotopy type of the data.

%%%%%%%%%%%%%%%%%%%%%%%%%%%%%%%%%%%%%%%%%%%%%%%%%%%%%%%%%%%%%%%%%%%%%%%%%%%
\paragraph{Contractibility for balls.}
Every non-empty convex set is contractible, which means it has
the homotopy type of a point.
The common intersection of two or more convex sets is either empty
or again convex and therefore contractible.
While primal Bregman balls are not necessarily convex, we show that 
their common intersections are contractible unless empty.
The reason for our interest in this property is
the Nerve Theorem \cite{Bor48,Ler45}, which asserts that the nerve
of a cover with said property has the same homotopy type
as the union of this cover.
\begin{result}[Contractibility Lemma for Balls]
  Let $\Bfun \colon \Domain \to \Rspace$ be of Legendre type,
  $X \subseteq \Domain$, and $r \geq 0$.
  Then $\bigcap_{x \in X} \Ball{\Bfun}{x}{r}$ is either empty or contractible.
\end{result}
\proof
  Recall the homeomorphism $h \colon \Domain \to \Domain^*$ obtained
  as a side-effect of applying the Legendre transform to $\Bfun$.
  It maps every primal Bregman ball in $\Domain$
  homeomorphically to a dual Bregman ball in $\Domain^*$, which is convex. 
  Similarly, it maps the common intersection of primal Bregman balls in $\Domain$
  homeomorphically to the common intersection of dual Bregman balls in $\Domain^*$:
  $h(\Xspace) = \Yspace$ in which
  $\Xspace = \bigcap_{x \in X} \Ball{\Bfun}{x}{r}$ and
  $\Yspace = \bigcap_{x \in X} \BallP{\Bfun^*}{x^*}{r}$.
  Since $\Xspace$ and $\Yspace$ are homeomorphic,
  they have the same homotopy type.
  Hence, either $\Xspace = \Yspace = \emptyset$ or
  $\Yspace$ is convex and $\Xspace$ is contractible.
\eop

%%%%%%%%%%%%%%%%%%%%%%%%%%%%%%%%%%%%%%%%%%%%%%%%%%%%%%%%%%%%%%%%%%%%%%%%%%%
\paragraph{\v{C}ech and Vietoris--Rips constructions for Bregman divergences.}
The contractibility of the common intersection suggests we take the
nerve of the Bregman balls.
Given a finite set $X \subseteq \Domain$ and $r \geq 0$,
we call the resulting simplicial complex
the \emph{\v{C}ech complex} of $X$ and $r$ associated with $\Bfun$.
Related to it is the \emph{Vietoris--Rips complex},
which is the clique complex of the $1$-skeleton of the \v{C}ech complex:
\begin{align}
  \Cech{\Bfun}{X}{r} &= \{ P \subseteq X \mid
     \bigcap_{p \in P} \Ball{\Bfun}{p}{r} \neq \emptyset \} ,
     \label{eqn:Cech} \\
  \Rips{\Bfun}{X}{r} &= \{ Q \subseteq X \mid \tbinom{Q}{2} \subseteq
     \Cech{\Bfun}{X}{r} \} .
     \label{eqn:VietorisRips}
\end{align}
In words, the Vietoris--Rips complex contains a simplex iff all its
edges belong to the \v{C}ech complex.
We note that for $\Bfun (x) = \norm{x}^2$,
\eqref{eqn:VietorisRips} translates to the usual Euclidean definition
of the Vietoris--Rips complex.
Increasing the radius from $0$ to $\infty$,
we get a filtration of \v{C}ech complexes and a filtration
of Vietoris--Rips complexes.
By construction, the \v{C}ech complex is contained in the Vietoris--Rips
complex for the same radius.
If we measure distance with the Euclidean metric, this relation extends to
\begin{align}
  \ECech{X}{r} \subseteq \ERips{X}{r} \subseteq \ECech{X}{\sqrt{2} r} .
  \label{eqn:interleaving}
\end{align}
Indeed, if all pairs in a set of $k+1$ balls of radius $r$ have
a non-empty common intersection, then increasing the radius to $\sqrt{2} r$
guarantees that the $k+1$ balls have a non-empty intersection.
This fact is often expressed by saying that the two filtrations
have a small interleaving distance if indexed logarithmically.

%%%%%%%%%%%%%%%%%%%%%%%%%%%%%%%%%%%%%%%%%%%%%%%%%%%%%%%%%%%%%%%%%%%%%%%%%%%
\paragraph{No interleaving.}
The interleaving property expressed in \eqref{eqn:interleaving}
extends to general metrics -- except that the constant factor is $2$
rather than $\sqrt{2}$ -- but not to general Bregman divergences.
To see that \eqref{eqn:interleaving} does not extend,
we give an example of $3$ points whose Bregman balls overlap pairwise
for a small radius but not triplewise until the radius is very large.

\begin{figure}[hbt]
 \centering {\includegraphics[width=2.8in]{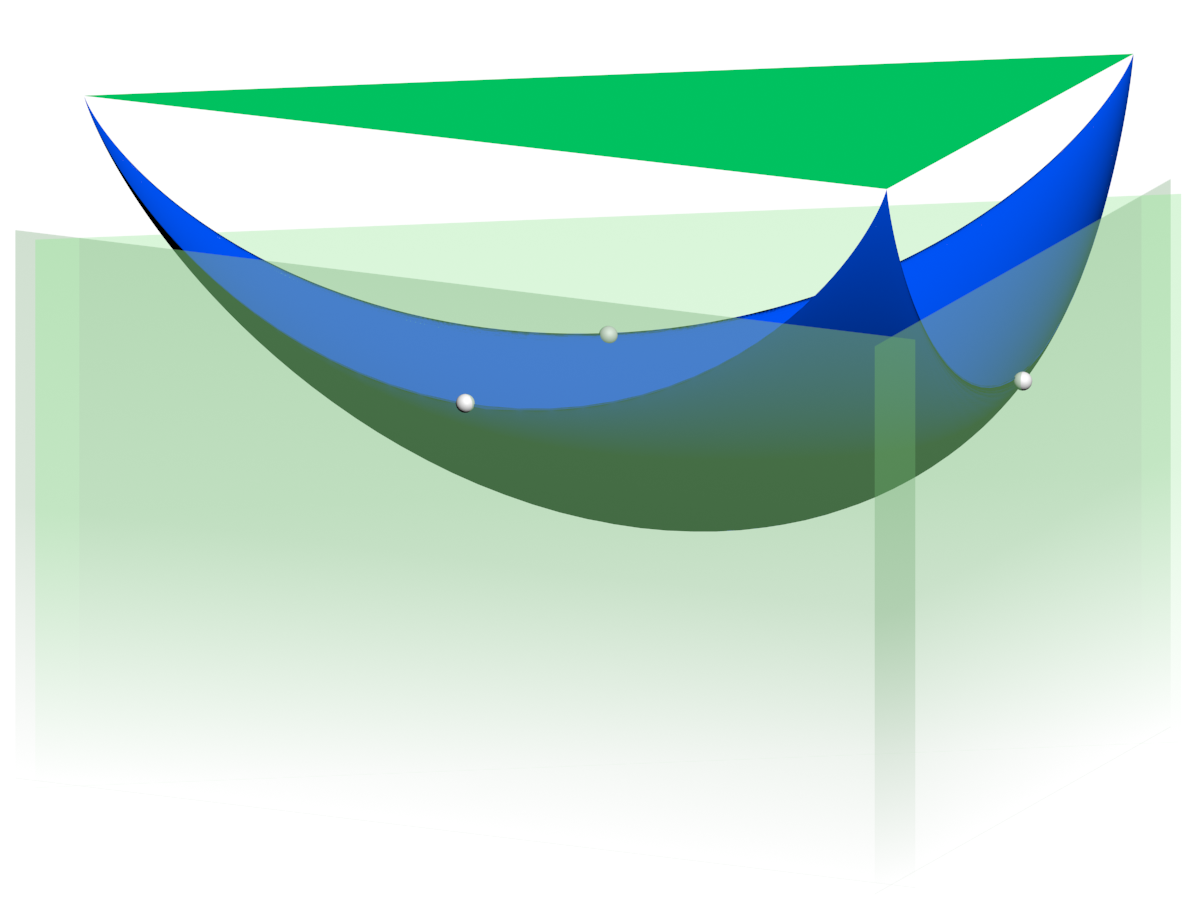}}
 \caption{Three points for which pairwise intersecting dual
   Kullback--Leibler balls centered at these points can be small,
   but triplewise intersecting such balls are necessarily large.}
 \label{fig:KL_with_planes}
\end{figure}

The example uses the exponential function defined on
the standard triangle, which we parametrize using barycentric coordinates.
For convenience, the explanation uses the conjugate function, which is the Shannon entropy;
that is: we look at dual balls in which distance is measured with the
Kullback--Leibler divergence.
Specifically, we use $\Bfun (x) = \sum_{i=1}^3 x_i \ln x_i$.
The barycentric coordinates are non-negative and satisfy
$\sum_{i=1}^3 x_i = 1$.
We therefore get the maximum value of $0$ at the three corners,
and the minimum of $- \ln 3$ at the center of the triangle;
see Figure \ref{fig:KL_with_planes}.
After some calculations, we get the squared length of the gradient
at $x$ as
$\tfrac{1}{3} [ (\ln x_1 - \ln x_2)^2
              + (\ln x_1 - \ln x_3)^2
              + (\ln x_2 - \ln x_3)^2 ]$.
It goes to infinity when $x$ approaches the boundary of the triangle.
We construct the example using points near the midpoints of the edges.
Choosing them in the interior of the triangle but close to the boundary,
the corresponding three tangent planes are as steep as we like.
Moving the planes upward, we get the dual balls as the vertical projections
of the parts of the graph of $\Bfun$ on or below the planes.
Moving the planes continuously, we let $r$ be the height above the
initial positions, and note that $r$ is also the radius of the dual balls.
Pairwise overlap between the balls starts when the three lines
at which the planes meet intersect the graph of $\Bfun$.
This happens at $r < \ln 3$.
Triplewise overlap starts when the point common to all three planes
passes through the graph of $\Bfun$.
This happens at a value of $r$ that we can make arbitrarily large.

%%%%%%%%%%%%%%%%%%%%%%%%%%%%%%%%%%%%%%%%%%%%%%%%%%%%%%%%%%%%%%%%%%%%%%%%%%%
\paragraph{Contractibility for Voronoi domains.}
\v{C}ech and Vietoris--Rips complexes can be high-dimensional
and of exponential size, even if the data lives in low dimensions.
To remedy this shortcoming, we use the Delaunay (or alpha) complex; see \cite{EdHa10,EdMu94}.
It is obtained by clipping the balls before taking the nerve.
We explain this by introducing the Voronoi domains of the generating
points as the clipping agents.
Letting $X \subseteq \Domain$ be finite, we define the
\emph{primal} and \emph{dual Voronoi domains} of $x \in X$ associated
with $\Bfun$ as the sets of points for which $x$ minimizes the Bregman
divergence to or from the point:
\begin{align}
  \Vdomain{\Bfun}{x}   &=  \{ a \in \Domain \mid
    \Bdist{\Bfun}{x}{a} \leq \Bdist{\Bfun}{y}{a}, \forall y \in X \} ; \\
  \VdomainP{\Bfun}{x}  &=  \{ a \in \Domain \mid
    \Bdist{\Bfun}{a}{x} \leq \Bdist{\Bfun}{a}{y}, \forall y \in X \} .
\end{align}
An intuitive construction of the primal domains grows the primal Bregman balls
around the points, stopping the growth at places where the balls meet.
Similarly, we get the dual Voronoi domains by growing dual Bregman balls.
Not surprisingly, the primal Voronoi domains are not necessarily convex,
and the dual Voronoi cells are convex.
To see the latter property, we recall that the dual ball
centered at $x$ is constructed
by translating the hyperplane that touches the graph of $\Bfun$ above $x$.
Specifically, $\Bdist{\Bfun}{a}{x}$ is the height at which the
hyperplane passes through the point $(a, \Bfun(a))$.
This implies that we can construct the dual Voronoi domains as follows:
\begin{itemize}\denselist
  \item  For each $x \in X$, consider the half-space of points
    in $\Rspace^n \times \Rspace$ on or above the hyperplane that touches
    the graph of $\Bfun$ at $(x, \Bfun(x))$.
  \item  Form the intersection of these half-spaces, which is a
    convex polyhedron.
    We call its boundary the \emph{upper envelope} of the hyperplanes,
    noting that it is the graph of a piecewise linear
    function from $\Rspace^n$ to $\Rspace$.
  \item  Project the upper envelope vertically onto $\Rspace^n$.
    Each dual Voronoi domain is the intersection of $\Domain$
    with the image of an $n$-dimensional face of the upper envelope.
\end{itemize}
We conclude that the dual Voronoi domains are convex
and use this property to show that the primal Voronoi domains
intersect contractibly.
\begin{result}[Contractibility Lemma for Voronoi Domains]
  Let $\Bfun \colon \Domain \to \Rspace$ be of Legendre type,
  and $X \subseteq \Domain$ finite.
  Then $\bigcap_{x \in X} \Vdomain{\Bfun}{x}$ is either empty or contractible.
\end{result}
The proof is similar to that of the Contractibility Lemma for Balls
and therefore omitted.
\ifskipme
\proof
  As in the proof of the Contractibility Lemma for Balls,
  we use the conjugate, $\Bfun^* \colon \Domain^* \to \Rspace$,
  and the homeomorphism, $h \colon \Domain \to \Domain^*$,
  obtained by applying the Legendre transform to $\Bfun$.
  By the Duality Lemma,
  we have $\Bdist{\Bfun}{x}{a} = \Bdist{\Bfun^*}{a^*}{x^*}$.
  Recall that the dual Voronoi domains associated with $\Bfun^*$ are convex.
  They are the homeomorphic images of the primal Voronoi domains
  associated with $\Bfun$.
  Similarly, $\Yspace = \bigcap_{x \in X} \VdomainP{\Bfun^*}{x^*}$
  is the homeomorphic image of $\Xspace = \bigcap_{x \in X} \Vdomain{\Bfun}{x}$.
  Hence, either $\Xspace = \Yspace = \emptyset$ 
  or $\Yspace$ is convex and $\Xspace$ is contractible.
\eop
\fi

%%%%%%%%%%%%%%%%%%%%%%%%%%%%%%%%%%%%%%%%%%%%%%%%%%%%%%%%%%%%%%%%%%%%
\paragraph{Delaunay construction for Bregman divergences.}
Taking the nerve of the primal Voronoi domains, we get
the \emph{Delaunay triangulation} of $X$ associated with $\Bfun$,
which we denote as $\DelTri{\Bfun}{X}$.
Further restricting the primal Voronoi domains by primal Bregman balls
of radius $r$,
we get the \emph{Delaunay complex} of $X$ and $r$ associated with $\Bfun$:
\begin{align}
  \!\!\Del{\Bfun}{X}{r}  &=  \{ P \!\subseteq\! X \mid
    \! \bigcap_{p \in P} \! \left[ \Ball{\Bfun}{p}{\!r} \cap \Vdomain{\Bfun}{p} \right]
    \neq \emptyset \} .
\end{align}
Assuming general position of the points in $X$,
the Delaunay triangulation is a simplicial complex of dimension at most $n$.
We will be explicit about what we mean by general position shortly.
Combining the proofs of the two Contractibility Lemmas,
we see that the common intersection of any set of clipped
primary balls is either empty or contractible.
This together with the Nerve Theorem implies that
$\Del{\Bfun}{X}{r}$ has the same homotopy type as $\Cech{\Bfun}{X}{r}$,
namely the homotopy type of the union of the Bregman balls that
define the two complexes.

%\clearpage
%%%%%%%%%%%%%%%%%%%%%%%%%%%%%%%%%%%%%%%%%%%%%%%%%%%%%%%%%%%%%%%%%%%%%%%%%%%
%%%%%%%%%%%%%%%%%%%%%%%%%%%%%%%%%%%%%%%%%%%%%%%%%%%%%%%%%%%%%%%%%%%%%%%%%%%
\section{Algorithms}
\label{sec4}
%%%%%%%%%%%%%%%%%%%%%%%%%%%%%%%%%%%%%%%%%%%%%%%%%%%%%%%%%%%%%%%%%%%%%%%%%%%
%%%%%%%%%%%%%%%%%%%%%%%%%%%%%%%%%%%%%%%%%%%%%%%%%%%%%%%%%%%%%%%%%%%%%%%%%%%

Recall that all three proximity complexes defined in Section \ref{sec3}
depend on a radius parameter. 
In this section, we give algorithms that compute the values of this
parameter beyond which the simplices belong to the complexes. 
By focusing on the resulting radius functions, 
we decouple the computation of the radius for each simplex
from the technicalities of constructing the actual simplicial complex. 
In particular, we show that the \v{C}ech complexes can be 
efficiently reconstructed from the Vietoris--Rips complexes, and the 
Delaunay complexes from the Delaunay triangulations.
We exploit a connection with discrete Morse theory to 
develop efficient algorithms.
%We begin by formalizing the values as functions.
%%%%%%%%%%%%%%%%%%%%%%%%%%%%%%%%%%%%%%%%%%%%%%%%%%%%%%%%%%%%%%%%%%%%%%%%%%%

\paragraph{Radius functions.}
Let $X \subseteq \Domain$ be finite, write $\Delta (X)$ for the simplex
whose vertices are the points in $X$,
and recall that $\DelTri{\Bfun}{X}$ is the Delaunay triangulation
of $X$ associated with $\Bfun$.
The \emph{\v{C}ech}, \emph{Vietoris--Rips}, and
\emph{Delaunay radius functions} associated with $\Bfun$,
\begin{align}
  \Cradius{\Bfun}     &\colon \Delta (X) \to \Rspace , \\
  \Rradius{\Bfun}\,   &\colon \Delta (X) \to \Rspace , \\
  \Dradius{\Bfun}\,\; &\colon \DelTri{\Bfun}{X} \to \Rspace ,
\end{align}
are defined such that
$P \in \Cech{\Bfun}{X}{r}$ iff $\Cradius{\Bfun} (P) \leq r$,
and similarly for Vietoris--Rips and for Delaunay.
By definition of the \v{C}ech complex,
$\Cradius{\Bfun} (P)$ is the minimum radius at which the primal
Bregman balls centered at the points of $P$ have a non-empty
common intersection.
We are interested in an equivalent characterization using dual
Bregman balls.
To this end, we say that a dual Bregman ball, $\BallPNo$,
\emph{includes} $P$ if $P \subseteq \BallPNo$, and we call $\BallPNo$ the
\emph{smallest including dual ball} if there is no other dual ball
that includes $P$ and has a smaller radius.
Because $\Bfun$ is strictly convex, the smallest including dual ball
of $P$ is unique; see Figure \ref{fig:itsaballs2},
which shows the smallest including dual Itakura--Saito ball of a pair of points.
%For later reference, we give a routine that take the center and radius of a
%dual ball as input and returns the data points in its interior.
% \begin{tabbing}
% m\=m\=m\=m\=m\=m\=n\=m\=\kill
% \> {\sf pointset} {\tt routine} {\sc In} $(q, r)$:                 \\*
% \> \> $Q = \emptyset$;                                             \\*
% \> \> {\tt forall} $a \in X$ {\tt do}                              \\*
% \> \> \> {\tt if} $\Bdist{\Bfun}{a}{q} \leq r$ {\tt then}
%                   $Q = Q \cup \{a\}$;                              \\*
% \> \> {\tt return} $Q$.
% \end{tabbing}
% The running time can be improved if we store the points in
% a spatial data structure;
% we refer to the Vietoris--Rips algorithm in \cite{Zom10},
% which tackles the same problem.
We call $\BallPNo$ \emph{empty} if no point of $X$ lies in its interior,
and we call it a \emph{circumball} of $P$
if all points of $P$ lie on its boundary.
We observe that a simplex $P \in \Delta (X)$ belongs to $\DelTri{\Bfun}{X}$
iff it has an empty dual circumball.
Because $\Bfun$ is strictly convex, the smallest empty dual circumball
of a simplex is either unique or does not exist.
The characterization of the radius functions in terms of dual balls
is strictly analogous to the Euclidean case studied in \cite{BaEd16}.
\begin{result}[Radius Function Lemma]
  Let $\Bfun \colon \Domain \to \Rspace$ be of Legendre type,
  $X \subseteq \Domain$ finite, and $\emptyset \neq P \subseteq X$.
  \begin{enumerate}\denselist
    \item[(i)]   $\Cradius{\Bfun} (P)$ is the radius of the smallest
      including dual ball of $P$, and
      $\Rradius{\Bfun} (P)$ is the maximum radius of the
      smallest including dual balls of the pairs in $P$. 
    \item[(ii)] Assuming $P \in \DelTri{\Bfun}{X}$,
      $\Dradius{\Bfun} (P)$ is the radius of the smallest
      empty dual circumball of $P$.
  \end{enumerate}
\end{result}

We omit the proof, which is not difficult.
Every circumball also includes, which implies
$\Rradius{\Bfun} (P) \leq \Cradius{\Bfun} (P) \leq \Dradius{\Bfun} (P)$
whenever the radius functions are defined.
Correspondingly,
$\Del{\Bfun}{X}{r} \subseteq \Cech{\Bfun}{X}{r} \subseteq \Rips{\Bfun}{X}{r}$
for every value of $r$.

%%%%%%%%%%%%%%%%%%%%%%%%%%%%%%%%%%%%%%%%%%%%%%%%%%%%%%%%%%%%%%%%%%%%%%%%%%%
\paragraph{General position.}
It is often convenient and sometimes necessary to assume that the points
in $X \subseteq \Domain$ are in general position,
for example when we require the Delaunay triangulation be a simplicial complex
in $\Rspace^n$.
Here is a notion that suffices for the purposes of this paper.
\begin{result}[Definition of General Position]
  Let $\Domain \subseteq \Rspace^n$ and $\Bfun \colon \Domain \to \Rspace$
  of Legendre type.
  A finite set $X \subseteq \Domain$ is in \emph{general position} with
  respect to $\Bfun$ if, for every $P \subseteq X$ of cardinality at most $n+1$,
  \begin{enumerate}\denselist
    \item[I.]  the points in $P$ are affinely independent,
    \item[II.] no point of $X \setminus P$ lies on the boundary
      of the smallest dual circumball of $P$.
  \end{enumerate}
\end{result}
Let $k = \dime{P}$.
Property I implies that $P$ has an $(n-k)$-parameter family of circumballs.
In particular, there is at least one circumball as long as $k \leq n$.
Property II implies that no two different simplices have the same
smallest dual circumball.
In particular, no two $n$-simplices in the Delaunay triangulation
have the same circumball.

%%%%%%%%%%%%%%%%%%%%%%%%%%%%%%%%%%%%%%%%%%%%%%%%%%%%%%%%%%%%%%%%%%%%%%%%%%%
\paragraph{Discrete Morse theory.}
For points in general position, two of the radius functions exhibit
a structural property that arises in the translation of Morse theoretic
ideas from the smooth category to the simplicial category.
Following \cite{BaEd16}, we extend the original formulation of
discrete Morse theory given by Forman \cite{For98}.
Letting $K$ be a simplicial complex, and $P, R \in K$ two simplices,
we write $P \leq R$ if $P$ is a face of $R$.
The \emph{interval} of simplices between $P$ and $R$ is
$[P,R] = \{ Q \in K \mid P \leq Q \leq R \}$.
We call $P$ the \emph{lower bound} and $R$ the \emph{upper bound}
of the interval.
A \emph{generalized discrete vector field} is a partition of $K$
into intervals.
We call it a \emph{generalized discrete gradient} if there
exists a function $f \colon K \to \Rspace$ such that
$f(P) \leq f(Q)$ whenever $P$ is a face of $Q$,
with equality iff $P$ and $Q$ belong to a common interval.
A function with this property is called
a \emph{generalized discrete Morse function}.
To get an intuitive feeling for this concept, consider the sequence
of sublevel sets of $f$.
Any two contiguous sublevel sets differ by one or more intervals,
and any two of these intervals are independent in the sense
that neither interval contains a face of a simplex in the other interval.
Indeed, this property characterizes generalized discrete Morse functions.
\begin{result}[GDMF Theorem]
  Let $\Bfun \colon \Domain \to \Rspace$ be of Legendre type  
  and let $X \subseteq \Domain$ be finite and in general position.
  Then $\Cradius{\Bfun} \colon \Delta (X) \to \Rspace$ and
  $\Dradius{\Bfun} \colon \DelTri{\Bfun}{X} \to \Rspace$
  are generalized discrete Morse functions.
\end{result}

We give the proof in Appendix \ref{appB}.
Observe that the Vietoris--Rips radius function is not a generalized
discrete Morse function.
The structural properties implied by the GDMF Theorem will be useful
in the design of algorithms that compute the radius functions.
The theorem should be compared with the analogous result in
the Euclidean case \cite{BaEd16}.
The arguments used there can be translated almost verbatim
to prove additional structural results for Bregman divergences.
Perhaps most importantly, they imply that the Wrap complex
of $\Bfun$ and $X$ is well defined -- see \cite{Ede03} for the
original paper on these complexes defined in $3$-dimensional Euclidean space --
and that the \v{C}ech complex collapses to the Delaunay complex
and further to the Wrap complex, all defined for the same radius.

%%%%%%%%%%%%%%%%%%%%%%%%%%%%%%%%%%%%%%%%%%%%%%%%%%%%%%%%%%%%%%%%%%%%%%%%%%%
\paragraph{Bregman circumball algorithm.}
Depending on how the function $\Bfun$ is represented,
there may be a numerical component to the algorithms
needed to find smallest including dual balls. 
Consider a $k$-simplex $Q \subseteq X$ with $0 \leq k \leq n$. 
Assuming general position, the affine hull of the points
$A = (a, \Bfun (a))$ with $a \in Q$ is a $k$-dimensional plane,
which we denote as $\QQQ$.
We are interested in the point $(q, \psi) \in \QQQ$ that maximizes
$\psi - \Bfun (q)$, the height above the graph of $\Bfun$.
The point $q$ is the center of the smallest dual circumball of $Q$,
and $\psi - \Bfun (q)$ is the radius. Interestingly, this observation 
implies  that the point of first intersection of two primal Bregman balls 
lies on a line joining their centers.
For later reference, we assume a routine that computes this point,
possibly using a standard numerical optimization method.

\begin{tabbing}
m\=m\=m\=m\=m\=m\=m\=m\=\kill
\> {\sf dualball} {\tt routine} {\sc CircumBall} $(Q)$:               \\*
\> \> let $\QQQ$ be the affine hull of the points $(a,\Bfun(a))$,
      $a \in Q$;                                                      \\*
\> \> find $(q, \psi) \in \QQQ$ maximizing $\psi - \Bfun (q)$;        \\*
\> \> {\tt return} $(q, \psi - \Bfun (q))$.
\end{tabbing}
This is an unconstrained $k$-dimensional convex optimization,
and $k$ is much smaller than $n$ for high dimensional data.
Indeed, the optimization can be performed in the space of affine coordinates
of the plane $\QQQ$.
Importantly, the Hessian is of dimension $k \times k$
and not $n \times n$, which would be prohibitive.
This allows us to use second-order quasi-Newton methods,
such as the fast BFGS algorithm~\cite{NoWr06}.

Note that the smallest dual circumball of $Q$ includes $Q$
but is not necessarily the smallest including dual ball.
However, the latter is necessarily the smallest dual circumball
of a face of $Q$.
Next, we show how the {\sc CircumBall} routine is
used to efficiently compute the radius functions.

%%%%%%%%%%%%%%%%%%%%%%%%%%%%%%%%%%%%%%%%%%%%%%%%%%%%%%%%%%%%%%%%%%%%%%%%%%%
\paragraph{\v{C}ech radius function algorithm.}
According to the Radius Function Lemma (i),
the value of a simplex, $Q \in \Delta (X)$, under the \v{C}ech radius function
is the radius of the smallest including dual ball of $Q$.
To compute this value, we visit the simplices in a particular sequence.
Recalling the GDMF Theorem,
we note that  the smallest including dual ball of a simplex $Q$
is the smallest dual circumball of the minimum face $P \subseteq Q$
in the same interval.
It is therefore opportune to traverse the simplices in the order
of increasing dimension.
Whenever the smallest dual circumball of a simplex $Q$ is
not the smallest including dual ball, we get $\Cradius{\Bfun} (Q)$
from one of its codimension $1$ faces.
We identify such a simplex $Q$ when we come across a face whose
smallest dual circumball includes $Q$,
and we mark $Q$ with the center and radius of this ball.
The following pseudocode computes the radius function of the
\v{C}ech complex restricted to the $k$-skeleton of $\Delta (X)$
for some nonnegative integer $k$:
\begin{tabbing}
m\=m\=m\=m\=m\=m\=m\=m\=\kill
\> {\tt for} $i = 0$ {\tt to} $k$ {\tt do}                              \\*
\> \> {\tt forall} $P \subseteq X$ {\tt with} $\dime{P} = i$ {\tt do}   \\*
\> \> \> {\tt if} $P$ unmarked {\tt then}
            $(p, r) = \mbox{\sc CircumBall} (P)$;                       \\*
\> \> \> {\tt forall} $a \in X$ with $\Bdist{F}{a}{p} < r $ {\tt do}
               mark $P \cup \{a\}$ with $(p, r)$.
\end{tabbing}
As in the Euclidean setting, the size of $\Delta (X)$ is exponential in
the size of $X$ so that the computations are feasible only
for reasonably small values of $k$ or small radius cut-offs. 
In practice, we would run the algorithm with a radius cut-off, or use 
an approximation strategy yielding a similar persistence diagram.

Observe the similarity to the standard algorithm for constructing the
$k$-skeleton of the Vietoris--Rips complex:
after adding all edges of length at most $2r$, we add simplices
of dimension $2$ and higher whenever possible.
Geometric considerations are thus restricted to edges and the rest of
the construction is combinatorial; see \cite{Zom10} for a fast implementation.
Our algorithm can be interpreted as constructing the \v{C}ech complex
from the Vietoris--Rips complex at the cost of at most one call
to {\sc CircumBall} per simplex.
This is more efficient than explicitly computing the smallest \emph{including}
dual ball for each simplex, even if we use fast randomized algorithms
as described in \cite{NiNo06,Wel91}.
Furthermore, the {\sc CircumBall} routine is only called
for the lower bounds of the intervals of the \v{C}ech radius function
or, equivalently, for each subcomplex in the resulting filtration.
The number of such intervals depends on the relative position of the
points in $X$ and not only on the cardinality.
Notwithstanding, the number of intervals is significantly smaller
than the number of simplices in the \v{C}ech complex.
This suggests that only a small overhead is needed to compute
the \v{C}ech from the Vietoris--Rips complexes.
Our preliminary experiments for the Kullback-Leibler divergence
support this claim; see Table~\ref{table:exp}.
Note that the number of calls to the {\sc CircumBall} routine
is between $\frac{1}{10}$ and $\frac{1}{3}$ of the number of simplices,
with an average between $6$ and $15$ function evaluations per call.

% Computing the full \v{C}ech complex for $18$ points in $\Rspace^{18}$ with
% $2^{18}-1=262,143$ simplices requires only $5,940$ calls to {\sc CircumBall}.
% For a more challenging example:
% the $3$-skeleton for $100$ points in $\Rspace^3$,
% with a radius cutoff of $0.15$ and $2,689,865$ simplices
% required $37,211$ calls to the {\sc CircumBall} routine.

% \begin{table}[h!]
% \centering
% \caption{Experimental evaluation. $S$, $O$, $F$ denote the number simplices, {\sc CircumBall} calls, and function evaluations in {\sc CircumBall}, respectively.}
% \label{table:exp}
% \begin{tabular}{llllll|l}
%  dim  &  0    &1    & 2    & 3    & 4    & $\Sigma$ \\
% S  &  4,000 &36,937 &143,475 &361,199 &677,077 &ca1,222,688  \\
% C  &  0 &37,003 &105697 &107,612 &33,310 &283,622  \\
% F  &  0 &126,660 &569,983 &778,140 &308,691 &1,783,474
% \end{tabular}
% \end{table}

\begin{table}[h!]
\centering
\caption{Experimental evaluation on three synthetic datasets:
({\bf{A}}) Full \v{C}ech complex with 20 points in $\Rspace^{20}$;
({\bf{B}}) $3$-skeleton with 256 points in $\Rspace^4$ and radius
cutoff $r=0.1$;
({\bf{C}}) $4$-skeleton  with 4,000 points in $\Rspace^4$ and radius cutoff $r=0.01$. 
}
\label{table:exp}
\begin{tabular}{l||rrr}
  & {\bf A} (20 pts) & {\bf B} (256 pts) & {\bf C} (4,000 pts)  \\
  \hline
  \hline
{\#}edges         & 190        & 7,715    & 36,937  \\
{\#}simplices        & 1,048,575    & 1,155,301  & 1,222,688 \\
{\#}calls to {\sc CircumBall}         & 104,030     & 346,475  & 283,622 \\
{\#}function evaluations in {\sc CircumBall} & 1,523,295    & 2,904,603 & 1,783,474  
\end{tabular}
\end{table}

%%%%%%%%%%%%%%%%%%%%%%%%%%%%%%%%%%%%%%%%%%%%%%%%%%%%%%%%%%%%%%%%%%%%%%%%%%%
\paragraph{Delaunay radius function algorithm.}
According to the Radius Function Lemma (ii),
the value of a simplex $Q \in \DelTri{\Bfun}{X}$ under the
Delaunay radius function is the radius of the smallest empty
dual Bregman circumball of $Q$.

%\vspace{-\baselineskip}
\begin{tabbing}
m\=m\=m\=m\=m\=m\=m\=m\=\kill
\> {\sf real} {\tt routine} {\sc DelaunayRadius} $(Q)$:             \\*
\> \> $(q, r) = \mbox{\sc CircumBall} (Q)$;                         \\*
\> \> {\tt forall} $a \in X \setminus Q$ {\tt do}                   \\*
\> \> \> {\tt if} $\Bdist{\Bfun}{a}{q} < r$ {\tt then}
            {\tt return} {\sf none};                                \\*
\> \> {\tt return} $r$.
\end{tabbing}
%\vspace{-\baselineskip}

The {\sc CircumBall} routine gives only the smallest dual circumball
of $Q$, and if it is not empty, then we have to get the value
of the Delaunay radius function from somewhere else.
According to the GDMF Theorem, we get the value from the
maximum simplex in the interval that contains $Q$.
It is therefore opportune to traverse the simplices of the Delaunay
triangulation in the order of decreasing dimension.
Whenever the smallest dual circumball of a simplex $Q$ is non-empty,
we get $\Dradius{\Bfun} (Q)$ from one of the simplices
that contain $Q$ as a codimension $1$ face.

As already observed in \cite{BNN10},
we can construct the full Delaunay triangulation, $\DelTri{\Bfun}{X}$, 
using existing algorithms for the Euclidean case. 
We get the Delaunay complexes as sublevel sets of the radius function.
Specifically, we first use the polarity transform to map the points
$(x, \Bfun (x))$ to the corresponding affine functions;
see Section \ref{sec2}.
We then get a geometric realization of $\DelTri{\Bfun}{X}$
from the vertical projection of the upper envelope of the
affine functions onto $\Rspace^n$,
which is a \emph{Euclidean weighted Voronoi diagram},
also known as \emph{power diagram} or \emph{Dirichlet tessellation}.
Its dual is the \emph{Euclidean weighted Delaunay triangulation},
also known as \emph{regular} or \emph{coherent triangulation}.
The data that defines these Euclidean diagrams are the points $x \in X$
with weights $\xi = \Bfun (x) - \norm{x}^2$.
Finally, after computing the radius function on all simplices in $\DelTri{\Bfun}{X}$, we get the Delaunay complexes as a filtration 
of this weighted Delaunay triangulation.
Interestingly, this is not necessarily the filtration
we obtain by simultaneously and uniformly increasing the weights
of the points.
%% Indeed, it is not difficult to see that given any filtration of
%% $\DelTri{\Bfun}{X}$, there exists a strictly convex function
%% that generates it.

%\clearpage
%%%%%%%%%%%%%%%%%%%%%%%%%%%%%%%%%%%%%%%%%%%%%%%%%%%%%%%%%%%%%%%%%%%%%%%%%%%
%%%%%%%%%%%%%%%%%%%%%%%%%%%%%%%%%%%%%%%%%%%%%%%%%%%%%%%%%%%%%%%%%%%%%%%%%%%
\section{Discussion}
\label{sec5}
%%%%%%%%%%%%%%%%%%%%%%%%%%%%%%%%%%%%%%%%%%%%%%%%%%%%%%%%%%%%%%%%%%%%%%%%%%%
%%%%%%%%%%%%%%%%%%%%%%%%%%%%%%%%%%%%%%%%%%%%%%%%%%%%%%%%%%%%%%%%%%%%%%%%%%%

The main contribution of this paper is the extension of the mathematical
and computational machinery of topological data analysis (TDA)
to applications in which distance is measured with a Bregman divergence.
This includes text and image data often compared with the
Kullback--Leibler divergence,
and speech and sound data often studied with the Itakura--Saito divergence.
It is our hope that the combination of Bregman divergences and
TDA technology will bring light into the generally difficult study
of high-dimensional data.
In support of this optimism, Rieck and Leitte \cite{RiLe15}
provide experimental evidence that good dimension reduction methods
preserve the persistent homology of the data.
With our extension to Bregman divergences, such experiments can
now be performed for a much wider spectrum of applications.
There are specific mathematical questions whose incomplete understanding
is currently an obstacle to progress in the direction suggested by this paper:
\begin{itemize}\denselist
  \item  A cornerstone of TDA is the stability of its persistence diagrams,
    as originally proved in \cite{CEH07}.
    How does the use of Bregman divergences affect the stability
    of the diagrams?
  \item  Related to the question of stability is the existence of
    sparse complexes and filtrations for data in Bregman spaces
    whose persistence diagrams are close to the ones we get for
    the \v{C}ech and Delaunay complexes.
\end{itemize}
Last but not least, we mention the urgent task to further study
the related algorithmic questions and to implement software that is fast,
can cope with large sets of data, and is easy to use also for non-specialists.

%\newpage
%%%%%%%%%%%%%%%%%%%%%%%%%%%%%%%%%%%%%%%%%%%%%%%%%%%%%%%%%%%%%%%%%%%%%%%%%%%
\subsection*{Acknowledgements}
{\small The authors thank \v{Z}iga Virk for discussions on the material
  presented in this paper.}
%%%%%%%%%%%%%%%%%%%%%%%%%%%%%%%%%%%%%%%%%%%%%%%%%%%%%%%%%%%%%%%%%%%%%%%%%%%

\appendix
\clearpage
%%%%%%%%%%%%%%%%%%%%%%%%%%%%%%%%%%%%%%%%%%%%%%%%%%%%%%%%%%%%%%%%%%%%%%%%%%%
%%%%%%%%%%%%%%%%%%%%%%%%%%%%%%%%%%%%%%%%%%%%%%%%%%%%%%%%%%%%%%%%%%%%%%%%%%%
\section{Polarity and Legendre Transform}
\label{appA}
%%%%%%%%%%%%%%%%%%%%%%%%%%%%%%%%%%%%%%%%%%%%%%%%%%%%%%%%%%%%%%%%%%%%%%%%%%%
%%%%%%%%%%%%%%%%%%%%%%%%%%%%%%%%%%%%%%%%%%%%%%%%%%%%%%%%%%%%%%%%%%%%%%%%%%%

In this appendix, we give further details how the polarity transform
amounts to the Legendre transform for functions of Legendre type.
Recall that the polarity maps a point
$C = (c, \gamma) \in \Rspace^n \times \Rspace$
to the function $C^* \colon \Rspace^n \to \Rspace$
defined by $C^* (x) = \scalprod{c}{x} - \gamma$,
and that it maps $C^*$ back to $(C^*)^* = C$.
Given a second point $S = (s, \sigma) \in \Rspace^n \times \Rspace$,
and the corresponding affine function $S^* (x) = \scalprod{s}{x} - \sigma$,
the transform preserves the difference between the values:
\begin{align}
  \sigma - C^* (s)  &=  \gamma - S^* (c) .
  \label{eqn:duality}
\end{align}
Indeed, both sides of the equation evaluate to
$\gamma + \sigma - \scalprod{c}{s}$.
To apply the polarity transform to $\Bfun$,
consider a point $A = (a, \Bfun(a))$ and note that the graph
of the affine function defined by
$B^*(x) = \Bfun(a) + \scalprod{\nabla \Bfun(a)}{x-a}$ is the hyperplane
that touches the graph of $\Bfun$ at $A$.
Let $P = (p, \Bfun(p))$ be another point on the graph of $\Bfun$
and $Q^*(x) = \Bfun(p) + \scalprod{\nabla \Bfun(p)}{x-p}$
the corresponding affine function.
To avoid potential confusion, we note that $B^*$ and $A^*$ are generally
different, and so are $Q^*$ and $P^*$.
Since $\Bfun$ is strictly convex, we have
\begin{align}
  \Bdist{\Bfun}{p}{a}  &=  \Bfun(p) - B^*(p)  \geq  0 , 
    \label{eqn:nonnegative1} \\
  \Bdist{\Bfun}{a}{p}  &=  \Bfun(a) - Q^*(a)  \geq  0 ,
    \label{eqn:nonnegative2}
\end{align}
with vanishing lefthand sides iff $a = p$; see Figure \ref{fig:conjugate}.
Applying the polarity transform,
we get two additional point/affine function pairs,
namely $B, A^*$ and $Q, P^*$.
We define $b = a^* = \nabla \Bfun(a)$ and $q = p^* = \nabla \Bfun(p)$
in $\Rspace^n$ and $\beta, \psi \in \Rspace$ such that
$B = (b, \beta)$ and $Q = (q, \psi)$; see again Figure \ref{fig:conjugate}.

Relating the two points with the two lines using \eqref{eqn:duality},
we get \eqref{eqn:dual1}, \eqref{eqn:dual2},
       \eqref{eqn:dual3}, \eqref{eqn:dual4}
by setting $C, S$ to $A, B$, to $P, Q$, to $P, B$, and to $A, Q$,
in this sequence:
\begin{align}
  \beta - A^* (b)  &=  \Bfun(a) - B^* (a),
    \label{eqn:dual1} \\
  \psi  - P^* (q)  &=  \Bfun(p) - Q^* (p),
    \label{eqn:dual2} \\
  \beta - P^* (b)  &=  \Bfun(p) - B^* (p),
    \label{eqn:dual3} \\
  \psi  - A^* (q)  &=  \Bfun(a) - Q^* (a).
    \label{eqn:dual4}
\end{align}
The two sides in \eqref{eqn:dual1} and in \eqref{eqn:dual2}
vanish by construction of $B^*$ and $Q^*$.
Using \eqref{eqn:nonnegative1} and \eqref{eqn:nonnegative2},
we see that the terms in \eqref{eqn:dual3} and \eqref{eqn:dual4}
are non-negative, and that they vanish iff $a = p$.
Their lefthand sides are the Bregman divergences between $b$ and $q$
under $\Bfun^*$, and their righthand sides can be rewritten using
the Duality Lemma:
\begin{align}
  \Bdist{\Bfun^*}{b}{q}  &=  \Bfun^* (b) - P^* (b) , \\
  \Bdist{\Bfun^*}{q}{b}  &=  \Bfun^* (q) - A^* (q) .
\end{align}
This provides the crucial inequalities that imply the required
properties of $\Bfun^*$, as enumerated in Section \ref{sec2}.

%%%%%%%%%%%%%%%%%%%%%%%%%%%%%%%%%%%%%%%%%%%%%%%%%%%%%%%%%%%%%%%%%%%%
\section{Discrete Morse Theory}
\label{appB}
%%%%%%%%%%%%%%%%%%%%%%%%%%%%%%%%%%%%%%%%%%%%%%%%%%%%%%%%%%%%%%%%%%%%

In this appendix, we present the proof of the GDMF Theorem.
Recall that this theorem claims that for
$\Bfun \colon \Domain \to \Rspace$ of Legendre type,
and $X \subseteq \Domain$ finite and in general position,
$\Cradius{\Bfun} \colon \Delta (X) \to \Rspace$ and
$\Dradius{\Bfun} \colon \DelTri{\Bfun}{X} \to \Rspace$
are generalized discrete Morse functions.
\proof
  We consider $\Cradius{\Bfun}$ first.
  Let $P \subseteq X$ be a $k$-simplex and consider two possibly different
  dual balls defined for $P$:
  the smallest including dual ball, $\BallP{\Bfun}{p_0}{r_0}$,
  and the smallest dual circumball, $\BallP{\Bfun}{p_1}{r_1}$.
  The first ball always exists,
  and by assumption of general position, the second ball exists iff $k \leq n$.
  We are interested in simplices for which the two balls are the same,
  which excludes simplices of dimension larger than $n$.
  They are the lower bounds of the intervals in the generalized
  discrete gradient \cite{BaEd16}.
  Let $P$ be such a simplex, and let $R$ be the set of points $x \in X$
  with $\Bdist{\Bfun}{x}{p_0} \leq r_0$.
  Clearly, $P \subseteq R$,
  and all simplices $P \subseteq Q \subseteq R$ have
  $\BallP{\Bfun}{p_0}{r_0}$ as the smallest including dual ball.
  All simplices in $[P, R]$ belong to $\Cech{\Bfun}{X}{r_0}$ but none
  of them belongs to $\Cech{\Bfun}{X}{r}$ with $r < r_0$.
  If $r_0 = r_1$ is unique for $P$, then this is the only difference
  between $\Cech{\Bfun}{X}{r_0}$ and its immediate predecessors.
  Else, the difference consists of two or more intervals.
  By assumption of general position, there are no face relations
  between the simplices in two different such intervals.
  It follows that $\Cradius{\Bfun}$ is a generalized discrete Morse function.

  We consider $\Dradius{\Bfun}$ second.
  The argument is similar, except that the relevant dual balls are different.
  Besides the smallest dual circumball of $P$, $\BallP{\Bfun}{p_1}{r_1}$,
  we also consider the smallest empty dual circumball,
  $\BallP{\Bfun}{p_2}{r_2}$.
  The latter exists iff $P$ belongs to the Delaunay triangulation of $X$.
  Again, we are interested in simplices for which the two balls are the same.
  They are the upper bounds of the intervals in the generalized
  discrete gradient \cite{BaEd16}.
  Let $P$ be such a simplex, and let $R$ be the smallest face of $P$
  such that the smallest containing dual ball of $R$ contains $P$.
  We note that in this case, $R$ has the same smallest empty dual circumball
  as $P$.
  Furthermore, $R \subseteq P$, and all simplices $R \subseteq Q \subseteq P$
  have the same smallest empty dual circumball.
  Hence, all simplices in $[R, P]$ belong to $\Del{\Bfun}{X}{r_1}$,
  and none of them belongs to $\Del{\Bfun}{X}{r}$ with $r < r_1$.
  If $r_1 = r_2$ is unique for $P$,
  then this is the only difference between $\Del{\Bfun}{X}{r_1}$
  and its immediate predecessors.
  Else, the difference consists of two or more intervals,
  and general position again implies that there are no face relations
  between the simplices in two different such intervals.
  It follows that $\Dradius{\Bfun}$ is a generalized discrete Morse function.
\eop

\newpage

%\newpage
%%%%%%%%%%%%%%%%%%%%%%%%%%%

\end{document}